# Free Will - a road less travelled in Quantum Information Processing

Ilyas Khan

St Edmunds College
University of Cambridge

Cambridge Quantum Computing

i.khan@jbs.cam.ac.uk

Feb 12, 2016

## Abstract

Conway and Kochen's Free Will Theorem is examined as an important foundational element within a new area of activity within computer science – developing protocols for quantum computing.

*Keywords: Free Will Theorem, Quantum Computing, Conway*

A few weeks before Christmas I went to listen to John Conway speak at the Isaac Newton Institute in Cambridge. The late afternoon talk took the form of a question and answer session moderated by Siobhan Roberts, who is Conway's biographer ( "Genius at Play - The curious mind of John Horton Conway" ). The initial invitations looked suspiciously like a book promotion (the biography noted above had only recently been published), but Conway has such a deep, committed and broad following among the Cambridge mathematics community that the event very quickly started to be touted as a seminar, and the large crowd that showed up meant that is was soon standing room only with the back walls pretty much two deep. The audience ranged from ex-colleagues and long-standing friends and collaborators of Conway, to a much bigger contingent of younger mathematicians and physicists for whom Conway was a largely mythical figure. Thankfully the book promotion part of the event was soon dispensed with, and I am sure that I was not alone in being relieved when Conway rescued his interlocutor (Siobhan Roberts) from being embarrassed by a rather banal q+a.

What might easily have become an awkward half hour that people had to endure before the promised cocktail party[1] very quickly became an enjoyable 90 minute seminar that very much lived up to the expectations that had drawn the crowd in the first place. That afternoon Conway resembled nothing if not an aging warrior of Greek mythology.

Conway, who has made Princeton his home for the past 30 years, is one of those very few living mathematicians who need absolutely no introduction whatsoever. His iconic Game of Life has inspired a whole sector of combinatorial game theory. There are a great many resources and literature on his Game of Life, but if you happen not to be familiar with the game or this area, then [this 1970 Scientific American Article](#) is still the best single introduction to an invention that has spawned a mini industry in computer science.

The great man was on fine form during the talk, and, not surprisingly, admitted to being a little bored by talking about the Game of Life. With a not so gentle dig at his biographer, Conway ditched any pretense at being a part of an obviously tired and choreographed q+a, and embraced his audience by settling down to talk about something 'more serious'. The result was a fascinating and at times utterly compelling extempore lecture about The Free Will Theorem ("FWT") that Conway developed

---

[1] It should be noted that the publishers, Bloomsbury did in fact put on a very lovely spread for cocktails

with his long time friend and Princeton colleague, Simon Kochen. Published in two parts, initially in 2006, and then again in 2009 (Free Will Theorem and Strong Free Will Theorem respectively) it is fair to say that Conway and Kochen's papers (I will refer to the joint authors as C+K) have provoked a deep, sharp, and long lasting response. With the passage of time the positives far outweigh the negatives but its still a little jarring to note some of the initial criticisms that were cloaked in the sort of language one associates with medieval prosecutors of the inquisition as opposed to normally conservative scientists, with some of the more trenchant early opposition generally being from physicists. And not only physicists (yes they still exist !) who recoil at FWT's rejection of the so called 'hidden variables' view of quantum mechanics. As we all now know, much of that early opposition has evaporated back from whence it came, and C+K's two papers now occupy a position of orthodoxy - something that I am sure provides Conway with a great deal of guilty pleasure[2] !

As a supplement to Bell's Theorem, and as a very clear successor to The Kochen-Specker theorem (itself an extension of Bell's Theorem) FWT's core assumptions sit comfortably at the modern heart of thinking within the theoretical physics community as it has developed since the early to mid 1960's. Conway himself, who readily and very happily admits to being an interloper among theoretical physicists (and thereby claims some protection behind Kochen who is anything but an interloper !) has not been shy of courting some friction in the way that he has framed his underlying objectives as being as much influenced by Philosophy as by Science. This is a contrived position, and says more about Conway's predilection for 'frissance' than it does about the paper. The fact is that the mathematical formalism that provides the essential scaffolding for the two papers meets the highest of rigorous standards. The papers are absolutely not philosophical treatises, and the 2009 additions were carried very prominently (rightly so) in the "Notices" of the American Mathematical Society.

---

[2] for a flavour of that mischievousness readers could do worse than watch some of the original Princeton videos of the formal presentation. They are available at https://www.youtube.com/watch?v=ftIllWczf5w

***"Do we really have free will, or, as a few determined folk maintain, is it all an illusion? We don't know, but will prove in this paper that if indeed there exist any experimenters with a modicum of free will, then elementary particles must have their own share of this valuable commodity"***

With these opening words in the 2006 paper, Conway and Kochen assert, through the course of 2 papers and about 4 years of work, that if we (ie human beings) have free will, then so do fundamental particles; that our decisions are not determined or influenced by prior conditions or prior history to any extent. The assertion, developed through three basic axioms - "fin" "spin" and "twin" initially appears to skirt with, but ultimately avoids anything remotely recursive. The three axioms that C+K detail in 2006 are updated in 2009:

**Spin** - particles have the 101 property (as per Kochen Specker)

**Twin** - assuming spin above, 2 particles possess total angular momentum of 0 (zero) and if the angular momentum of one particle is measured as a, then the angular momentum of the other particle is -a (minus a)

**Fin** - this 2006 axiom which reflects the speed of light as the upper bound for any velocity for the transmission of information is replaced in 2009 by **Min** - where the two experimenters are space-like separated and both A and B can equally be 'first' in observing or measuring their states.

From these simple bases the track to a conclusion is straightforward[3]. In the enhanced 2009 paper, C+K reiterate the p(home) and p(away) entanglement thought experiment and conclude with the statement:

***"The Free Will Theorem thus shows that any such theory, even if it involves a stochastic element, must walk the fine line of predicting that for certain interactions the wave function collapses to some eigen function of the Hamiltonian, without being able to specify which eigen function this is"***

Any reasonable analysis around FWT and the direct linkages with causality in quantum mechanics took the original conversations very quickly towards the so-called "open problem" or the free will loop hole in Bell's theorem. Given the fact that we now know that large parts of that debate have been 'de-fanged' I will skip that part of the conversation. For those interested in some of the specific experimental advances during the 5 years from FWT 2009 I refer by way of example to developments at MIT's astronomy lab is described [here](here).

It might be useful at this stage to pause and consider Free Will as a concept and remind ourselves of how to frame a possible definition. Martin Gardner (see more on him below) provides as good a summary of Free Will as I have come across. He says:
"***Free will is neither fate, nor chance. In some unfathomable way it partakes of both***".

Many of us of an older vintage (by which I mean more than 40 years old) are likely to be drawn, when thinking about any discussion of FWT, to the well known 'Newcomb Paradox'. There are many different

---

[3] It should be noted that in a review article in AMS Notice Goldstein and Tausk have suggested that only the probability distribution implied in Min should be independent of each other in the case cited of A and B. Kochen is on record as requiring that both the probability distribution and the outcome should be independent. I have listed the Goldstein et al article in the references below

descriptions of this well known principle of choice, but one of my favourite online versions is part of a video series on 'famous' maths problems by Norman Wildberger (link here). If you get a chance I do recommend watching it. Wildberger brings a mathematical focus to the conversation that is extremely well put together.

Wildberger refers in his video to the same Martin Gardner that I refer to above, and of course the quote that I use is from the very same book by Gardner that Wildberger reads from. It is "The Colossal Book of Mathematics" and the quote that I have taken above is from the introduction to the section about Newcomb's Paradox. This is a link to this amazing book. Gardner is one of those mathematicians who is universally praised by those who knew him or learnt from him. Additionally he was a huge influence on a whole generation who read his regular column in Scientific American that examined mathematical problems in a popular but non-condescending manner. Gardner's book is actually a collection of those columns, and it is never very far from my desk.

Back now to NWT. I was talking about the Newcomb Paradox and how Wildberger creates a very cool little mathematical exposition and then solution to the challenge. Wildberger's solution is situated partly in a traditional probability and statistical part of the maths universe and partly in the mathematical physics that describe expectation values in quantum mechanics. His equation representing his 'solution' for the paradox is:

$$P(a) + P(b) \geq 1 + 1/k$$

*where P(a) is the probability of the Newcomb oracle being correct in her 'guess' when the chooser chooses option A (ie the larger prize) and P(b) is the probability of the Newcomb oracle being correct in her guess when the chooser chooses option B (ie both prizes). "k" is simply the*

*result of the amount represented by the money given in the larger prize (in the newcomb paradox this is normally described as $1million), being divided by the amount guaranteed in the smaller prize ($1000).*

If you dont know what Newcomb's Paradox states, and you have not yet watched the video, I recommend reading [this](#) very short summary.

The undying appeal about Newcomb's Paradox is that pretty much anyone who studies the problem ends up believing, and emphatically believing, that they are sure of the right answer, that everyone and anyone with the opposite view is silly and obviously a few shillings short of a pound, and that that the "right" thing to do is obvious. Over a period of time research has shown that opinions end up being virtually 50:50 between those who think that Option A is the ONLY logical and correct decision. Equally, there are those (50% in fact) who think that the only possible right answer is Option B, and that anyone who thinks otherwise.....etc etc

Wildberger's formula is itself a straightforward reduction of what he calls "expected value". As the terminology suggests there is a link to the approach taken in measuring observables in quantum mechanics, and this is necessary not because the paradox is one of quantum mechanics, but because it does not specify, with any degree of computational accuracy just how reliable the Oracle will be in guessing your choice, and pretty much the same mathematical approach that works with expectation values in quantum states also works here.

We know (we are told in the rules and guidelines to the paradox) that in the overwhelming majority, but not all cases, the Oracle is correct. This means that even if the rate of accuracy is 99% there will still be that odd occasion when the Oracle is wrong. This uncertainty, meaning we just don't know what the accuracy rate is , along with the fact the oracle's mind, once made up, cannot be changed (ie you can make your choice at the very last minute and it will not affect what the Oracle has already decided) has meant that the paradox has been extremely useful in highlighting core issues around the problem of free will not merely by philosophers, but also physicists and mathematicians. In many ways the paradox is a worthy vehicle for such discussions. Even when allowing for the fact that a participant in the game will amend her actions depending upon how many times she is allowed to play (it becomes very easy to choose ONLY the big box if you know you can play the game many times), the implications are clear - maybe even when you think you are making a free choice, you are not actually free of some degree of causality. Or so we think !

I should point out at this stage that the degree of uncertainty that is embedded in the Newcomb Paradox ought not to be taken as an analogue of quantum entanglement. In any truly entangled state we can know all there is to know about a system without knowing anything about the individual state of a particle within the entangled system. The analogy between uncertainty and knowledge (of the oracle and by the oracle) in Newcomb's paradox, and quantum entanglement is a slippery path if taken very far. FWT is not a riff on entangled states. As we shall see the proof of FWT does indeed utilise the laws of quantum physics as they relate to uncertainty and entanglement, but the debate around free will and the different references that are made to Newcomb when talking about free will stop short of making any judgements about those links.

Two important formal pieces of work on Free Will that appeared in the aftermath of C+K's work do however raise the linkage between Free Will and quantum mechanics and computing in a much more credible manner, and one of the papers actually makes ample use and reference to Newcomb's Paradox. I refer to Seth Lloyd's entertaining and engrossing scholarly work "[A Turing Test for Free Will](#)" (2013) and Scott Aaronson's equally engaging lecture on [Free Will](#) (2014) that is also included as a later chapter in his book "[Quantum Computing since Democritus](#)".

Both Seth Lloyd and Scott Aaronson have the incomparable advantage of having an interest and expertise in quantum mechanics and quantum computing - the two areas that are of greatest interest to me when considering C+K's FWT. Lloyd's approach to the subject is to neither argue in favour or against Free Will, but to arrive at the conclusion that we all (that is we human beings) tend towards a position where we can't know or have any knowledge about our own personal decisions more fully than the point at which the decisions are actually made. In other words even if we try to anticipate our own decisions, we will fail.

The only thing we can do is to actually make those decisions, not try and pre-form them. His paper is based on a Turing Test that proves this point.

***"The inscrutable nature of our choices when we exercise free will is a close analogue of the halting problem: once we set a train of thought in motion, we do not know whether it will lead anywhere at all. Even if it does lead somewhere, we do not know where that somewhere is until we get there. Ironically, it is customary to assign our own unpredictable behaviour and that of others to irrationality: were we to behave rationally, we reason, the world would be more predictable. In fact, it is just when we behave rationally, moving logically like a computer from step to step, that our behaviour becomes provably unpredictable. Rationality combines with the capacity of self-reference to make our actions intrinsically paradoxical and uncertain"***

The Halting problem and Turing's development of Godel's incompleteness theorem have long been a final destination or stopping point for mathematicians (and that tiny handful of philosophers who are equipped to do understand Turing or Godel's formalism) that have considered the problem of Free Will from a classical point of view. I mean by the use of the word 'classical' that they have avoided the uncertainty inherent in quantum physics and use an approach that is clearly traditional, hence 'classical', but not random.

In fact Lloyd (and Aaronson) dispenses very quickly with the concept of randomness (and in doing so they align themselves clearly with C+K) in order to develop a more rigorous construction for exhibiting a pre-disposition for Free Will.

Aaronson's step by step defusing of the case for randomness as a by product of the Newcomb paradox is actually clearer than C+K's approach (I will speak more on that below). In the case of Lloyd however, the Turing Test that he designs, allows individuals to inhabit the duality of accepting a deterministic universe, but one where we are individually convinced of our free will. His paper, in his own words sets out to:

***"show that any Turing simulatable decision-making process leads us to intrinsically unpredictable decisions, even if the underlying process is completely deterministic".***

Ultimately Lloyd's position is that it would take a deterministic protocol longer to "determine" the decisions about to be taken by a "decider" than it takes for a decider to make her decision. Moreover it takes longer (and is more complicated) for us to try and decide our own decisions ahead of them being made, leading to the strongly held view that we all believe in our power of free will and we believe we are endowed with at least some element of freedom. Lloyd stops short of expressing a personal decision on the matter, and in my view that is the only real criticism that can be leveled at the paper since the whole structure and approach that he adopts seems to imply a stance in favour of C+K's FWT. If that is the case, then why not simply acknowledge the fact ?

Aaronson is less shy, and, as one might expect, far more directly emphatic in stating his opinion that Free Will is the axiomatic tenet of a universe in the post Bell's experimental era. Aaronson takes the results of the Kochen Specker theorem and also C+K's FWT, as merely useful extensions and confirmations of Bell, rather than separate or stand alone theorems[4]. He takes pretty much the same view as Wildberger in consigning Newcomb's paradox into being a challenge that ultimately says very little about Free Will. Where Wildberger then proves his point by 'solving' the paradox, Aaronson develops the view that Newcomb's paradox is not really very challenging once one assumes that one has free will, and the idea that the Oracle of supreme being who knows your decision can influence your decision in a directly causal or deterministic manner is untenable. Aaronson does not arrive at this decision with an explicit formula as Wildberger did, but gets to the same destination by careful deduction:

***"One reason I like this Newcomb's Paradox is that it gets at a connection between "free will" and the inability to predict future behavior. Inability to predict the future behavior of an entity doesn't seem sufficient for free will, but it does seem somehow necessary. If we had some box, and if without looking inside this box, we could predict what the box was going to output, then we would probably agree among ourselves that the box doesn't have free will".***

As we have seen, in contrast to Aaronson, Seth Lloyd stops at this point of the debate, choosing to expand upon his self designed "Turing test". Aaronson however pushes on beyond this initial boundary and

---

[4] In neither of C+K's two papers, 2006 or 2009 do they address Von Neumann's view of dispersion free states, but it is my view that Bell's rather peremptory dismissal of Von Neumann in this context and subsequent work have dispensed with any requirement to comment on this related issue. Bell's collection of essays "Speakable and Unspeakable in Quantum Mechanics" contains a thorough review of this topic.

addresses C+K directly. Reading the Aaronson paper akin to the experience of realising that the summit we have just climbed is a false peak, and the real trophy lies further ahead. There is still some hard climbing to do. I am reminded of Piet Hein's quaint rhyme:

***"A bit beyond perception's reach/I sometimes believe I see/That life is two locked boxes, each/Containing the other's key"***

As I note above, Aaronson appears to place C+K as merely another confirmation of Bell's theorem and therefore similar to the [Kochen-Speker Theorem.](#) Interestingly however, Aaronson's articulation of C+K's initial assumptions provide a sort of ground state for the FWT.

These assumptions are: the freedom to choose a quantum state; the ability to perform experiments on the Alice and Bob states in separate spacetime frames such that both Alice and Bob could be "first'; and finally that information cannot travel quicker than lightspeed.

Aaronson arrives at his conclusion:

***"The measurement outcomes can't have been determined in advance, even probabilistically; the universe must "make them up on the fly" after seeing how Alice and Bob set their detectors".***

Shortly after I was a member of the audience for Conway's memorable talk in Cambridge, I travelled over to the west coast and spent some time with the Artificial Intelligence team within Google at their headquarters just off Venice Beach in LA. Like all who visit that facility, I am constrained by an NDA in talking about what is going on. However in their bid to establish "Quantum Supremacy" the team, led by Hartmut Neven, talks not in terms of decades but in a timetable that is the technology equivalent of tomorrow. For the avoidance of doubt, the "tomorrow" that I refer to is the timeline for building and operating a universal quantum computer. What is utterly fascinating is that whether we examine Google's efforts, or those of other combatants in the race such as Microsoft or IBM, a C+K type articulation and view of Free Will represents the basis upon which most of the departures from classical computing are being designed.

In the short time since the publication of the second FWT paper in 2009, the fundamental assumptions that C+K established in their paper have already been utilised in what I would describe as a 'real world' and commercial problem. And of course we all know that during the early part of this period (ie from about 2009 until about the middle of 2014), progress towards building a quantum computer had yet to really heat up. That period that might be described as 'heating up" did not start until the middle of 2014 when engineering advances started to be used by large organisations with the budget and the infrastructural resource to build and maintain long term programmes. The engineering advances I refer to by way of example, are the same ones that have led to progress at the Large Hedron Collider, and more recently to the ground breaking experiments at LIGO (gravitational waves of course !).

More specifically, the use of FWT related precepts from C+K's paper that I refer to, is the use in quantum information science of randomness[5] generators to deliver something called an Einstein Certified random number ("ECR number). In its simplest form an ECR number carries with it the guarantee that is absolutely random. Short of faster than light travel or some form of spooky time travel, an ECR number is about as certainly random as we could possibly devise, and for all kinds of commercial reasons any protocols that can be adapted for quantum computers in this area will be of very great importance.

---

[5] Tumulka 2006 comments meaningfully on randomness. In a paper that is otherwise trenchantly hostile to C+K the short section on randomness is very good as it contains a useful exposition of the views of C+K in consigning stochastic processes as ultimately deterministic

In his book "Quantum Computing since Democritus" Aaronson very briefly picks up on this development and explains how Bells and C+K's Alice and Bob experiments on entangled particles result in a true rolling of the dice **"on the fly"**. It is that last statement that brings to the fore the importance of Free Will in Quantum Information Sciences.

As regular readers of my occasional articles will know, I have been rather more optimistic about the emergence of a truly quantum computational processor than most, and within the very narrow confines of working out how we might usefully benefit from such machines, (narrow in the sense that I don't think we will get to grips with the question until there are real machines accessible to more than a handful of theoretical physicists hidden away in large corporates or government defence and security funded programmes), true randomness has very quickly become very important.

For those interested in looking a little more deeply at the subject, my favourite practical paper on the subject of random number generation is one that I have only recently encountered, though it was written in the tail end of 2013 (and therefore published after Aaronson's book). Written by a clutch of researchers, ([link here to arxiv source](#)) the paper covers ground that provides a concrete setting for precisely those aspects of C+K's FWT that matter the most. The abstract to the paper concludes with the comment:

**" The protocol involves four devices, can amplify any non-deterministic source into a fully random source, tolerates a constant rate of error, and has its correctness based solely on the assumption of no-signaling between the devices".**

A little later, in the main body of the paper the authors state:

***"The existence of non-local quantum correlations violating Bell inequalities already suggests that device-independent randomness amplification could be achieved. However, the violation of Bell inequalities requires that the measurements be performed in a random manner, independent of the system upon which they are performed".***

Sound familiar ? Where have we heard that phrase before ?
That *'modicum of free will'* that C+K assume in FWT is the foundation from which this attempt at true randomness is being generated.

In Aaronson's book on Quantum Computing, he creates a basis for a class of problems that can be efficiently solved using a quantum as opposed to a classical computer. You will recall that BPP or the Bounded Error Probabilistic Polynomial Time class of problems refers to those problems that are capable of being solved on a classical computer. The class of problems that can be solved using a Quantum Computer is known as BQP or Bounded Error Quantum Polynomial Time problems.

When defining the parameters that will contain such problems, Aaronson (and others) identify 4 items or qualities that need to be present. 1) Initialization; 2) Transformations; 3) Measurement; and 4) Uniformity. The first three, viz initialization (initialized states or qubits), transformations (the state of superposition essential for any operators to be effected), and measurement (normalised observables) can be simulated but not perfected unless one assumes that Bell's theorem works all the time, everytime and everywhere, and therefore that all particles (qubits) are endowed with Free Will. Note, Free Will, and not mere randomness.

C+K may not have anticipated this particular direction of travel for their work from over 10 years ago. Their thought experiment, posited as FWT, is now a core principle that could, in my view, drive the utilisation of whatever power quantum computing is going to provide to its users, and the same basis (for example) that provides true randomness will also provide true quantum effectiveness. When Feynman and others talked about quantum computers finally bringing us to the point where we can compute using nature's own methods, they assumed not only that quantum entanglement works (after all it is experimentally one of the most successful of modern theories), but that the results are neither deterministic nor random.